\documentclass{aa}  

\usepackage{graphicx}
\usepackage{txfonts}
\usepackage{hyperref}

\usepackage{xcolor} 
\usepackage{txfonts} 
\usepackage{rotating} 
\usepackage{url}  
\usepackage{mathtools, amsmath, amssymb, amsfonts} 
\usepackage{ccicons} 
\usepackage{mathabx}
\usepackage{multirow}
\usepackage{comment}
\usepackage{placeins}
\usepackage{subcaption}

\DeclareUnicodeCharacter{2212}{-}

\graphicspath{{Figures/}}

\defcitealias{2021ApJ...910...51K}{Kom21}
\defcitealias{2021MNRAS.508.3611P}{Pic21}
\defcitealias{2023A&A...673A.139A}{A23}

\begin{document} 

    \title{The evolution of the flux-size relationship in protoplanetary discs by viscous evolution and radial pebble drift
    }

   \author{J. Appelgren\inst{1},
          A. Johansen\inst{1,2},
          M. Lambrechts\inst{2},
          J. Jørgensen\inst{3}, 
          N. van der Marel\inst{4},
          N. Ohashi\inst{5},
          and
          J. Tobin\inst{6}
          }

   \institute{
    \inst{1} Lund Observatory, Division of Astrophysics, Department of Physics, Lund University,  Box 43, 22100 Lund, Sweden \\
    \inst{2} Center for Star and Planet Formation, Globe Institute, University of Copenhagen, Øster Voldgade 5-7, 1350 Copenhagen, Denmark \\
    email: Anders.Johansen@sund.ku.dk  \\
    \inst{3} Centre for Star and Planet Formation, Niels Bohr Institute \& Natural History Museum of Denmark, University of Copenhagen, Øster Voldgade 5–7, 1350 Copenhagen K., Denmark \\
    \inst{4} Leiden Observatory, Leiden University, P.O. Box 9531, NL-2300 RA Leiden, the Netherlands \\    
    \inst{5} Academia Sinica Institute of Astronomy \& Astrophysics, 11F of Astronomy-Mathematics Building, AS/NTU, No.1, Sec. 4, Roosevelt Rd, Taipei 10617, Taiwan \\
    \inst{6} National Radio Astronomy Observatory, 520 Edgemont Rd., Charlottesville, VA 22903, USA.
    }

   \date{Received ; accepted }

    \authorrunning{J. Appelgren et al.}

\abstract{In this paper we study the evolution of radiative fluxes, flux radii and observable dust masses in protoplanetary discs, in order to understand how these depend on the angular momentum budget and on the assumed heat sources. We use a model that includes the formation and viscous evolution of protoplanetary gas discs, together with the growth and radial drift of the dust component. We find that we are best able to match the observed fluxes and radii of class 0/I discs when we assume (i) an initial total angular momentum budget corresponding to a centrifugal radius of 40\,au around solar-like stars, and (ii) inefficient viscous heating. Fluxes and radii of class II discs appear consistent with disc models with angular momentum budgets equivalent to centrifugal radii of both 40\,au or 10\,au for solar like stars, and with models where viscous heating occurs at either full efficiency or at reduced efficiency. During the first $\sim$$0.5$\,Myr of their evolution discs are generally optically thick at $\lambda=1.3$\,mm. However, after this discs are optically thin at mm-wavelengths, supporting standard means of dust mass estimates. Using a disc population synthesis model, we then show that the evolution of the cumulative evolution of the observable dust masses agrees well with that observed in young star forming clusters of different ages.}

 


   \maketitle
%

\section{Introduction}

The evolution of dust in protoplanetary discs is a necessary component to explain planet formation. Pebble accretion is, in principle, capable of forming both giant planets and rocky planets \citep{2012A&A...544A..32L, 2017A&A...604A...1O, 2019MNRAS.484.1574T, 2021SciA....7..444J, 2024A&A...682A..43G}, but requires an influx of drifting pebbles. The magnitude of the pebble flux determines if the forming planets are super-Earths or terrestrial \citep{2019A&A...627A..83L}. The concentration of pebbles into rings due to the presence of pressure bumps can then also lead to the rapid and efficient formation of planetesimals. Such planetesimal rings could facilitate the rapid formation of rocky planets and super-Earths from pure planetesimal accretion \citep{2023NatAs...7..330B}. Understanding how dust evolution depends on stellar mass and disc mass is therefore necessary to know what type of planetary systems are created around which stars.
 
Observational measurements of dust masses in protoplanetary discs frequently rely on the assumption that dust discs are optically thin \citep[e.g.][]{2016ApJ...828...46A}. However, if the dust disc emission is partially, or completely optically thick, a large amount of dust can remain hidden below the optically thick surface. This possible problem has been studied by a number of papers \citep[e.g.][]{2012A&A...540A...6R, 2018ApJ...868...39G, 2019AJ....157..144B}, showing that dust masses may be underestimated by a factor of several up to an order of magnitude \citep{2022A&A...668A.175L}. Further, if dust discs are optically thick, the scattering of light on dust particles can reduce the emission, making an optically thick disc appear optically thin \citep{2019ApJ...877L..18Z}.

Accurate estimates of disc masses also rely on reasonable assumptions about disc temperature and opacities. A temperature of 20\,K is often chosen for class II discs, based on SED modelling of protoplanetary discs from the Taurus region \citep{2005ApJ...631.1134A}. This modelling assumed that dust discs are 100\,au in size. However, we now know that typical dust disc sizes are $\sim$$40$\,au \citep{2020ApJ...895..126H}, and even this lower number is strongly affected by a large fraction of discs that have only upper limits on their size. The dust opacity depends on both particle size and observing wavelength. The assumed disc opacity is selected from a dust opacity models that take into account the dust's maximum particle size, composition, porosity and size distribution \citep{2016A&A...586A.103W, 2018ApJ...869L..45B}. Interestingly, porous particles have a flatter dependence on the maximum particle size \citep{2014A&A...568A..42K}. This is in contrast to compact particles, where the opacity obtains its maximum value approximately where the particle size matches the observing wavelength. 

Determining the size of a protoplanetary disc is not straightforward either, as the disc gradually fades into the environmental conditions of the giant molecular cloud. A practical solution to measuring disc sizes self-consistently is therefore to use the flux radius, defined as     the radius containing some fraction of the total flux \citep{2017ApJ...845...44T}. 
A given mm flux radius is determined by integrating the intensity profile until the desired percentage of the flux is found. However, there is no consensus on which percentage is most useful. Common choices include 68\%, 90\%, and 95\% \citep{2019A&A...626L...2F, 2020ApJ...895..126H, 2022ApJ...931....6L}.

Here, we will investigate the theoretical relationship between the flux and the flux radius of protoplanetary discs.
\citet{2017ApJ...845...44T} first reported a notable correlation between mm flux and flux radius using SMA observations of discs in mainly Taurus and Ophiuchus.
This approximately linear scaling of the mm emission with emitting surface was then found to hold also in ALMA observations of discs in the Lupus star-forming region \citep{2018ApJ...865..157A,2021MNRAS.506.2804T}.
Such a relation appears to be broadly consistent with inwards pebble drift, although the observed sample shows significant scatter \citep{2019MNRAS.486L..63R}.

In this work we will take an evolutionary perspective, modeling disc flux and radius throughout the formation and evolution of the gas disc, addressing stellar and disc diversity (Sec.\,\ref{sec:model}). 
We find in this paper, that discs with weak viscous heating and high angular momentum are best able to reproduce the observed disc fluxes and flux radii (Sec.\,\ref{sec:results}).
In second part of this work, we explore if this model is also consistent with the important constraint placed by the observed dust mass reduction with cluster age (Sec.\,\ref{sec:res:popsynth}). This aspect of the study builds upon earlier disc population synthesis work \cite{2023A&A...673A.139A}, hereafter \citetalias{2023A&A...673A.139A}. 
In this way, 
we show that discs undergoing radial drift are optically thin for the majority of their lifetimes. The cumulative distribution of the optically thin dust masses agree well with dust masses observed in nearby star-forming regions. We discuss model limitations (Sect.\,\ref{sec:limits}),  before presenting 
our conclusions (Sect.\,\ref{sec:conclusions}).

\section{Model and data} \label{sec:model}

\subsection{Protoplanetary disc model}

\begin{table}[t]
\centering
\caption{Parameters used for the two disc models investigated in this paper}
\begin{tabular}{cccc} 
    \hline \hline 
    Model                   & $\alpha_\mathrm{vh}$  & $R_1$     & $\alpha_\nu$ \\ 
                            &                       & (au)        &              \\ \hline
    $\texttt{mod-a-3-r40}$  & $10^{-3}$             & 40        &    $10^{-2}$ \\
    $\texttt{mod-a-2-r10}$  & $10^{-2}$             & 10        &    $10^{-2}$ \\ 
    \hline
\label{tab:mod_params}
\end{tabular}
\end{table}

We used the protoplanetary disc model from \citetalias{2023A&A...673A.139A}, with some modifications. Here, we provide a brief explanation of the model, for more details, we refer to \citetalias{2023A&A...673A.139A}. The gas disc model includes the formation of a protoplanetary disc from a collapsing Bonnor-Ebert sphere following \cite{2013ApJ...770...71T} and viscous evolution using the classical $\alpha$-disc model \citep{1973A&A....24..337S, 1981ARA&A..19..137P}. The dust disc evolves via advection with the gas and through radial drift \citep{1977MNRAS.180...57W}. We trace only the largest dust particles that contain a large majority of the dust mass. The dust particle size is set by the fragmentation limit \citep{2012A&A...539A.148B}, assuming a fragmentation velocity of $v_\mathrm{f} = 1$\,m/s and a turbulent diffusion coefficient of $\alpha_t = 10^{-4}$. We separate the viscous $\alpha$-parameter,  $\alpha_\nu$, which drives angular momentum redistribution, from the turbulent stirring parameter $\alpha_t$, that affects the fragmentation limit. 

In order to investigate how the flux-size relationship of protoplanetary discs depends on both the angular momentum of the molecular cloud core and viscous heating in the disc, we present two variations of the disc model: one model features higher angular momentum with viscous heating at reduced efficiency, while the other model has lower angular momentum with viscous heating at full efficiency.

We set the angular momentum of the cloud core by scaling the centrifugal radius as $R_\mathrm{c} = R_1 (M_\mathrm{core}/M_\odot)\ \mathrm{au}$. In the high angular momentum model, $R_1 = 40$\,au, and in the low angular momentum model $R_1=10$\,au. The low-angular-momentum model uses the same scaling as in \citetalias{2023A&A...673A.139A}. Less efficient viscous heating is achieved by using $\alpha_\mathrm{vh} = 10^{-3}$ when calculating the viscous heating rate, compared to $\alpha_\mathrm{vh} = 10^{-2}$ in the model with viscous heating at full efficiency. We note that we keep the angular transport coefficient at a high level of $\alpha_\nu = 10^{-2}$, assuming that 90\% of the angular transport results from disc winds \citep{2013ApJ...769...76B}. {With the exception of the population synthesis, explained in Sect. \ref{sec:res:popsynth}, we do not include photoevaporation in these models, as we are focused on how the disc size evolves under radial drift and viscous evolution. The model parameters used in each model are shown in Table \ref{tab:mod_params}.

\begin{figure}[t]
    \centering
    \includegraphics[width=\hsize]{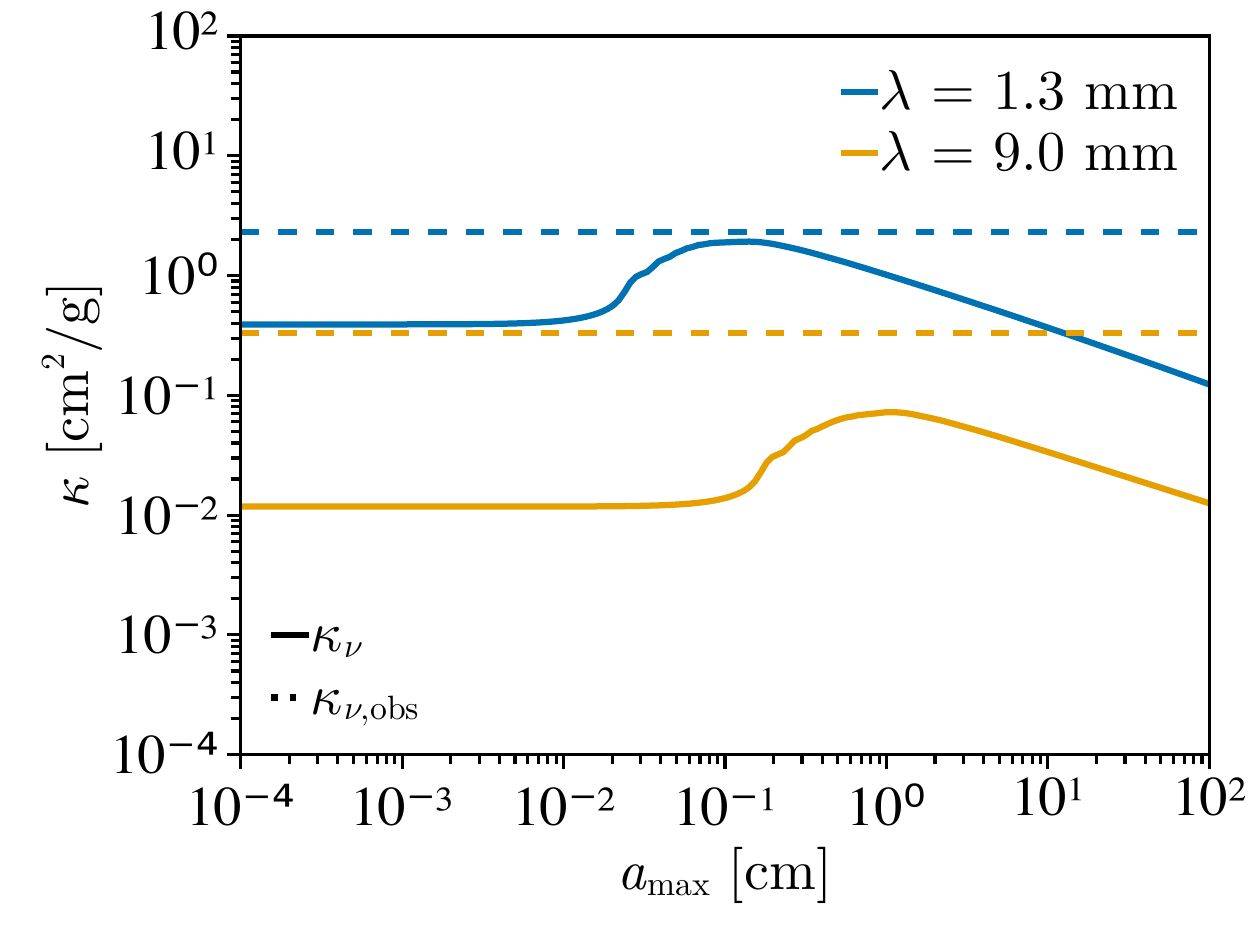}
    \caption{The opacity as a function of maximum particle size, at an observing wavelength of 1.3\,mm (blue) and 9.0\,mm (yellow) calculated from the \cite{2018ApJ...869L..45B} opacity model. The dashed lines represent the assumed disc opacity used in \cite{2016ApJ...828...46A}.}
    \label{fig:mod:opacity_amax}
\end{figure}

\subsection{Observable and derived quantities} \label{sec:model:Obs_Md}

\begin{figure*}[t]
    \centering
    \includegraphics[width=\hsize]{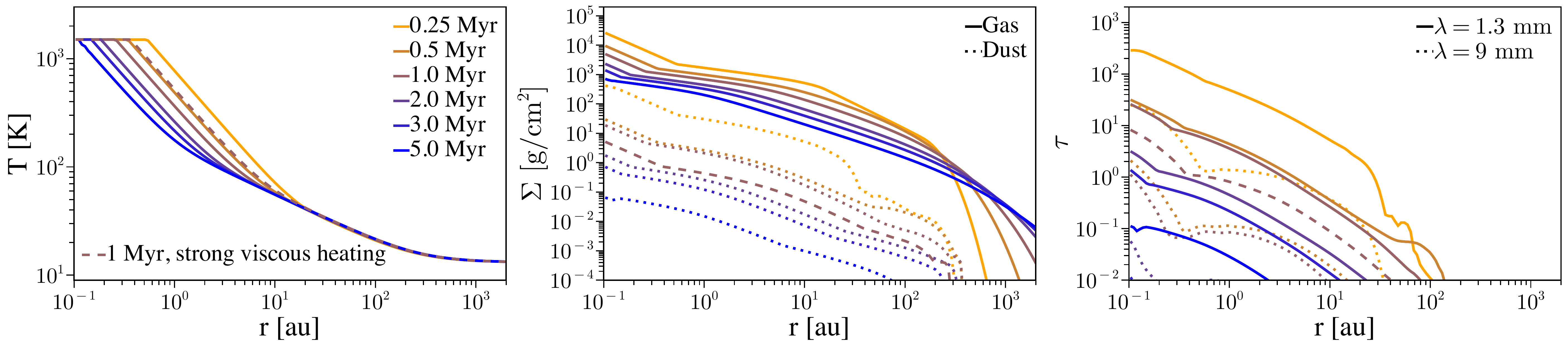}
    \caption{Evolution of the temperature (left panel), the gas surface density and dust surface density (solid and dotted lines, middle panel) and optical depth (right panel) at $\lambda = 1.3$\,mm (solid lines) and $\lambda = 9.0$\,mm (dotted lines)  as a function of orbital radius at different disc ages. During the build-up of the disc and shortly after (0.25 Myr), the disc is optically thick out to $\sim$$40$\,au. Up to 1\,Myr the disc remains optically thick within $\sim$$10$\,au, and after this the disc is optically thin except for the inner $\sim$$1$\,au. The dashed line shows a model with strong viscous heating and low angular momentum at 1 Myr. The temperature corresponds in this case well to the 0.5 Myr case with weak viscous heating.}
    \label{fig:surf_dens_opt_depth}
\end{figure*}

\begin{figure}
    \centering
    \includegraphics[width=\hsize]{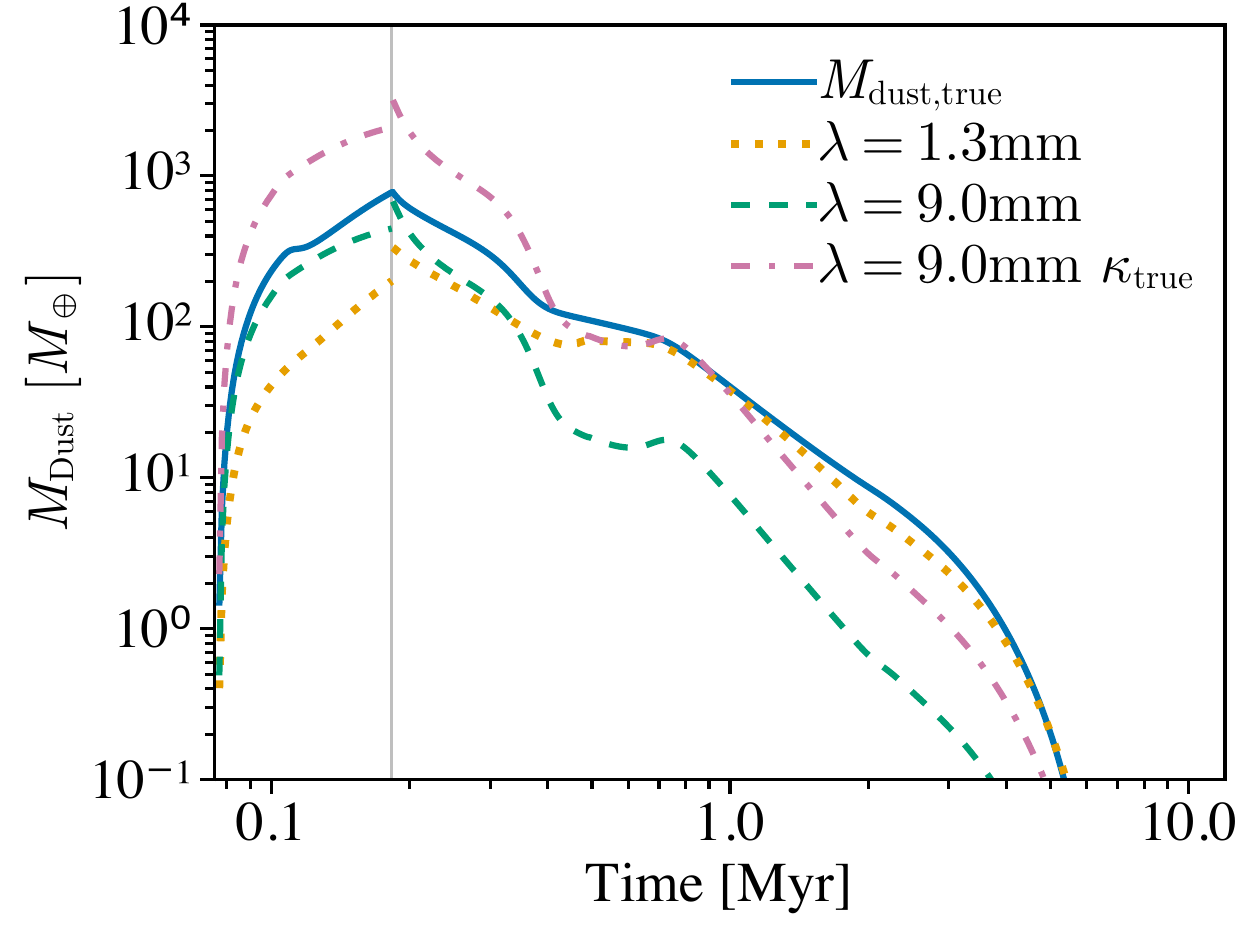}
    \caption{The true dust mass of the model (blue), the optically thin dust mass at an observing wavelength of 1.3\,mm (yellow dotted) and 9\,mm (green dashed), as a function of time. The observations assume a disc temperature of 20\,K during the formation of the disc (< 0.18\,Myr, grey vertical line) and 30\,K at later times. The pink dashed-dotted line shows the 9\,mm, optically thin dust mass, where the opacity used to calculate this in Eq. \eqref{eq:MdustObs} has been set to the opacity that the \cite{2018ApJ...869L..45B} model produces at a wavelength of 9\,mm and at a maximum particle size of 9\,mm, i.e. the same opacity that is used to calculate the disc continuum emission. This is to highlight how the resulting optically thin dust mass can be affected by a difference in between actual disc opacity, and the scaling law (Eq. \ref{eq:kappa_obs}) that is often used to determine the opacity in observational surveys of protoplanetary discs.}
    \label{fig:res:Mdust_obs}
\end{figure}

Under the assumption that the discs are optically thin, the dust mass can be calculated from the observed flux with the following equation \citep{1983QJRAS..24..267H},
\begin{align}
    M_\mathrm{Dust,obs} = \dfrac{F_\nu d^2}{\kappa_{\nu,\mathrm{obs}} B_\nu(T_\mathrm{disc})}\,. \label{eq:MdustObs}
\end{align}
Here, $F_\nu$ is the flux received at Earth's distance to the disc, $d$ is the distance to the disc from the observer, $\kappa_{\nu,\mathrm{obs}}$ is the assumed disc opacity, and $T_\mathrm{disc}$ is the assumed disc temperature. Observational surveys commonly make use of Eq. \eqref{eq:MdustObs} to estimate the dust mass in protoplanetary discs. When using Eq. \eqref{eq:MdustObs}, a characteristic temperature and opacity of the disc must be chosen. 

A common choice for the dust temperature is 20\,K for class II objects, based on fits of spectral energy distribution models to protoplanetary discs \citep{2005ApJ...631.1134A, 2016ApJ...828...46A}. Class 0 and I objects are hotter than class II objects and a temperature of 30\,K is used, based on model SEDs of young stellar objects \citep{2003ApJ...598.1079W, 2020A&A...640A..19T}. We note that more accurate estimates of the dust mass taking into account the effect of the  stellar luminosity and the disc size on the disc temperature exist \citep[e.g.][]{2020ApJ...890..130T, 2022ApJ...934...95S}. When reporting the optically thin dust masses via Eq. \eqref{eq:MdustObs}, we will assume a disc temperature of $T_\mathrm{disc} = 30\ \mathrm{K}$ when the disc is forming (class 0/I stage) and $T_\mathrm{disc} = 20\ \mathrm{K}$ after the disc has finished forming (class II stage) for consistency with the typical choices as made in  \cite{2016ApJ...828...46A, 2020A&A...640A..19T}, as we are comparing our model results with their observational work.
 
We use Eq. \eqref{eq:MdustObs} to calculate what we refer to as the optically thin dust mass in our disc model. When calculating this, scale the assumed disc opacity, $\kappa_{\nu,\mathrm{obs}}$, with wavelength by
\begin{align}
    \kappa_{\nu,\mathrm{obs}} = 10\ \mathrm{cm}^2 \mathrm{g}^{-1} \left(\dfrac{\nu}{1000\ \mathrm{GHz}} \right) \label{eq:kappa_obs}
\end{align}
as in \cite{2016ApJ...828...46A}. 

In order to calculate the total emission from our synthetic protoplanetary discs, we use 
\begin{align}
    I_\nu(T) = B_\nu(T)\left(1 - e^{-\tau_\nu}\right). \label{eq:Intensity}
\end{align}
Here, $B_\nu$ is the Planck function and $\tau_\nu$ is the optical depth along a vertical column. The optical depth along the line of sight can be determined from the particle size dependent opacity at the observed wavelength as 
\begin{align}
    \tau_\nu = \dfrac{\kappa_\nu(a_\mathrm{max}) \Sigma_\mathrm{dust}}{\cos{i}},
\end{align} 
where $\Sigma_\mathrm{dust}$ is the vertically integrated dust surface density, and $i$ is the inclination of the disc relative to the observer. To determine the actual opacity of the disc, $\kappa_\nu(a_\mathrm{max})$, we use opacity tables from the model of \cite{2018ApJ...869L..45B}, assuming a particle size distribution of $n(a)\propto a^{-3.5}$. Figure \ref{fig:mod:opacity_amax} shows the resulting opacity as a function of maximum particle size, assuming a particle size distribution with a slope of -3.5, at an observing wavelength of 1.3\,mm and at 9.0\,mm.

From the intensity, the observed flux of a disc can be calculated using the following equation
\begin{align}
    F_\nu = \dfrac{2\pi\cos{i}}{d^2}\int_{R_\mathrm{in}}^{R_\mathrm{out}}\! R I_\nu(T) \mathrm{d}R \label{eq:Flux}
\end{align}
where $d$ is the distance from the observer to the source which we set to 100\,pc, and $R_\mathrm{in}$ and $R_\mathrm{out}$ are the inner and outer edge of the disc. We assume that all discs in our model are observed face-on, and therefore set $i = 0^\circ$

\begin{figure*}
    \centering
    \includegraphics[width=\hsize]{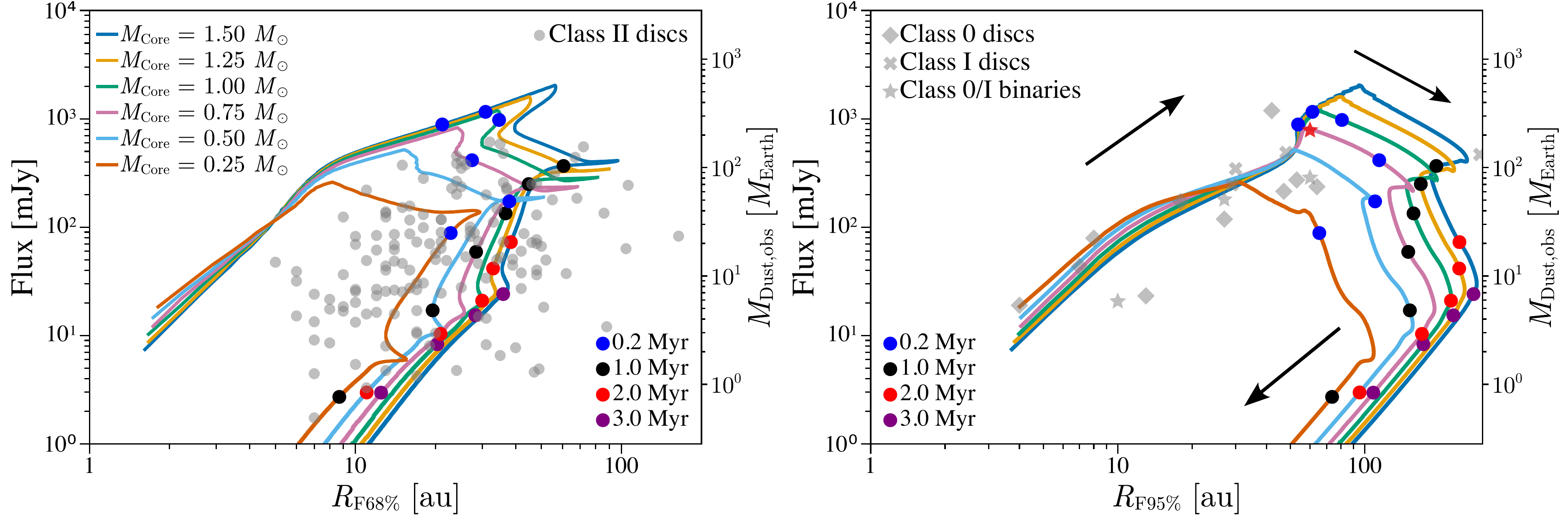}
    \caption{The total emitted flux at $\lambda = 1.3$\,mm versus the flux radius, comparing to observed class II disc in the left panel and class 0/I discs in the right panel (gray symbols). The left and right panel show the 68\% and 95\% flux radius because the observational samples we compare to report 68\% flux radii for the class II discs, and the class 0/I discs report the 95\% flux radius. The class II sample is taken from \cite{2016ApJ...831..125P, 2018ApJ...869L..41A, 2018ApJ...859...21A, 2019MNRAS.482..698C, 2019ApJ...872..158A, 2021AJ....162...28V} and the class 0/I discs from \cite{2023ApJ...951....8O}. We show resolved discs with measured sizes and do not include those with upper limits on their size. The arrows show the direction of the temporal evolution in this plane, and the coloured dots show the ages of the model discs at a few times of their evolution. The red star indicates the disc R CrA IRS7B-a, which has indications of active viscous heating \citep{2024ApJ...964...24T}.
    }
    \label{fig:Flux_Rflux_lowvisc}
\end{figure*}

We determine the flux radius of the discs, $R_\mathrm{flux}$, by integrating the intensity profile from Eq. \eqref{eq:Intensity} radially until the desired percentage of the total flux has been found as
\begin{align}
    \dfrac{\int_{r_\mathrm{in}}^{R_\mathrm{flux}}\!I_\nu\mathrm{d}r}{F_\nu} = P_\mathrm{flux},
\end{align}
where $P_\mathrm{flux}$ is the chosen fraction of the total flux, either 0.68 or 0.95. A better comparison with observations could be achieved by sampling the intensity profile at a spatial resolution similar to what has been achieved in surveys of discs. Integrating this profile to find the flux radius would more accurately mimic the size finding procedure of observational surveys. However, assuming that this procedure traces the true intensity accurately, the differences between the two methods should be small. Therefore, we opt to measure our flux radii from the actual model.


\section{Results} \label{sec:results}

In this section, we examine the evolution of the optical depth and optically thin dust mass of a disc formed from a 1\,$M_\odot$ cloud core. With this we can understand the evolution of the population synthesis model.

\subsection{Evolution of a single disc} \label{subsec:singledisc}

For the longest part of its evolution, the dust disc is optically thin at mm wavelengths in the regions where most of the mass is located. Figure \ref{fig:surf_dens_opt_depth} shows the evolution of the disc temperature in the left panel, of the gas and dust surface density as a function of orbital radius in the middle panel, and of the optical depth as a function of orbital radius in the right panel. Looking at the right panel, the disc is optically thick as it is forming and shortly thereafter (0.25 Myr snapshot). Up to 1 Myr, the disc remains optically thick in the inner $\sim$$10$\,au. However, looking at the middle panel of Fig. \ref{fig:surf_dens_opt_depth} shows that the dust disc extends to $\sim$$300$\,au, and as such much of the mass resides in the outer optically thin regions of the discs. As the disc evolves further, it becomes optically thin throughout the disc.
We note that $t=0$ in these simulations is the time when the collapse of the molecular cloud core begins. Because it takes time for the centrifugal radius to reach the inner edge of the simulated grid (0.1\,au), the time when the disc obtains a resolved size is $t\approx$$8\times 10^4$\,yr.

\subsubsection{Observing at $\lambda = 1.3$\,mm}

The optically thin dust mass as a function of time is shown in Fig. \ref{fig:res:Mdust_obs}. During the first 0.25\,Myr the disc is optically thick at $\lambda = 1.3$\,mm, due to the very high dust surface densities that are present as the disc is built up. The resulting high optical depth causes the true dust mass at $\lambda = 1.3$\,mm to be underestimated by a factor of between 2-3 as shown by the yellow dashed curve. Once the disc is older than about 0.5\,Myr, the disc is optically thin and the true dust mass is only underestimated by at most factor of 1.5. This remaining inaccuracy of the optically thin dust mass is caused by a discrepancy between the assumed representative disc temperature and opacity and their actual values. An error in the temperature can cause either an overestimation or underestimation of the disc mass if the assumed temperature is too low or too high respectively. An error in the assumed opacity will have a similar effect. Assuming a too high opacity will result in underestimating the disc mass, and a too low opacity will overestimate the disc mass. From Fig. \ref{fig:mod:opacity_amax} we can see that the assumed disc opacity at $\lambda = 1.3$\,mm (blue dotted line) traces the opacity used to calculate the fluxes (blue solid line) accurately at particle sizes of $\sim$$1$\,mm. At particle sizes smaller or larger than this, the assumed opacity is higher than the true disc opacity. Therefore, with the opacity models we use, a discrepancy in opacity always leads to underestimating the dust mass.

\begin{figure*}[t]
    \centering
    \includegraphics[width=\hsize]{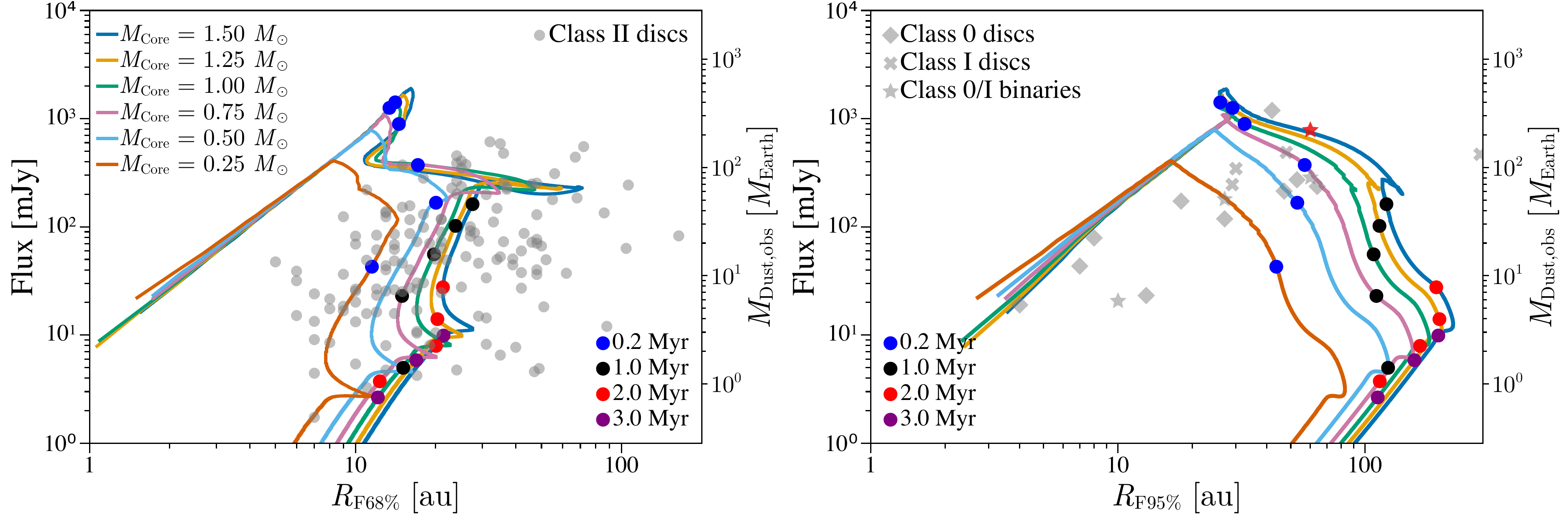}
    \caption{Emitted flux $\lambda = 1.3$\,mm as a function of the 68\% (left) and 95\% (right) flux radius, similar to Fig. \ref{fig:Flux_Rflux_lowvisc}, for low-angular momentum discs with strong viscous heating. These discs show smaller radii at a given flux than the high-angular momentum, weakly viscously heated discs from Fig. \ref{fig:Flux_Rflux_lowvisc}. At the same time, they produce too high fluxes and too low radii when compared to the class 0/I sample. The red star shows the disc around R CrA IRS7B-a, which has signs of active viscous heating \citep{2024ApJ...964...24T}.}
    \label{fig:Flux_Rflux}
\end{figure*}

\subsubsection{Observing at $\lambda = 9$\,mm}

At $\lambda = 9$\,mm the optical depth of the disc is significantly lower than at $\lambda = 1.3$\,mm. As a consequence, the optically thin disc mass at $\lambda = 9$\,mm traces the true disc mass more accurately during the first 0.25\,Myr. The disc mass is only slightly underestimated. As the disc evolves, however, unlike in the $\lambda = 1.3$\,mm case, the degree of underestimation grows rather than shrinks. This surprising behaviour occurs because the particle size becomes smaller with time. When the gas surface density decreases over time, the fragmentation limit occurs at smaller particle sizes, and thus the maximum particle size decreases over time. In turn, the difference between the true and assumed disc opacity at $\lambda = 9$\,mm becomes larger, and the inaccuracy in the estimated dust mass grows. If our model included a particle growth mechanisms that could allow dust grains to grow past the fragmentation barrier, such as dust trapping in pressure bumps with weak turbulence, this behaviour might not be observed.

When calculating the optically thin dust mass at $\lambda = 9$\,mm, we scaled the assumed disc opacity using Eq. \eqref{eq:kappa_obs}. The resulting opacity at 9\,mm from this scaling law is higher than produced by the \cite{2018ApJ...869L..45B} opacity model. Part of the explanation why we find that the dust mass is underestimated lies in this difference in opacity. If we reduce this source of error by fixing the assumed disc opacity to that of the \cite{2018ApJ...869L..45B} at $\lambda = 9$\,mm and maximum particle size of 9\,mm, the resulting mass estimate is shown as the pink line in Fig. \ref{fig:res:Mdust_obs}. The result is now that the dust mass is overestimated during the first 0.4\,Myr and underestimated at later times. The overestimation during the formation of the disc happens because the true disc temperature is higher than the assumed 30\,K used when calculating the optically thin dust mass.

\subsection{The flux-radius relationship} \label{sec:res:Flux_Rflux}

Estimating the dust mass from the measured disc flux relies on assumptions about disc optical depth. If the disc is optically thin, accurately estimating the mass with Eq. \eqref{eq:MdustObs} still relies on correctly choosing the representative disc temperature and representative disc opacity. As these are selected from SED modelling of discs and opacity models, respectively, the resulting disc mass is model-dependent. Uncertainty in the interpretation of our results can be vastly reduced by comparing quantities that have less model dependence. We take a step in this direction by correlating the total emitted disc flux to the flux radius. The total flux that the modelled disc is calculated to emit will still depend on the assumed opacity model and on the method used to calculate the flux. But, by examining the flux directly, we do not have to choose a representative disc opacity and a representative disc temperature. Our observable, the flux, will depend directly on the disc temperature and dust distribution. Observational measurements on the flux radius depend on the assumed intensity profile. However, the 68\% radius appears to not be very sensitive to the assumed profile \citep{2020A&A...633A.114S}.

Figure \ref{fig:Flux_Rflux_lowvisc} shows the evolution of the flux at $\lambda = 1.3$\,mm as a function of the flux radius. The colours show discs created from cloud cores of different masses. The left panel shows a comparison with class II discs, and the right panel shows a comparison with class 0/I discs. We show the 68\% flux radius in the left panel and the 95\% flux radius in the right panel, as the disc sizes reported from observational surveys we compare to give either the 68\% or 95\% flux radii. The right y-axis shows the dust mass equivalent to the flux on the left y-axis, using the optically thin Eq. \eqref{eq:MdustObs} and assuming a disc temperature of 20\,K. The coloured dots show the ages of the disc at a few stages in their evolution.

The protoplanetary discs are initially small with low fluxes. As the discs continue to grow, they move upward in the left branch, which is the epoch where the disc forms and is still embedded in its cloud core. This branch thus physically represents the observational class 0 and I phases. When the disc leaves the formation phase, the flux radius continues to increase while the flux decreases slowly. The expansion of the flux radius occurs due to the radial expansion of the disc, caused by the outward transport of angular momentum. In addition to this, the flux radius effectively expands because the inner disc cools down and contributes less to the total flux. After this, both the flux and flux radius decrease as the disc is drained of dust by radial drift. This right branch represents the observational class II objects, since these discs are no longer embedded.  

\begin{figure*}[t]
    \centering
    \includegraphics[width=\hsize]{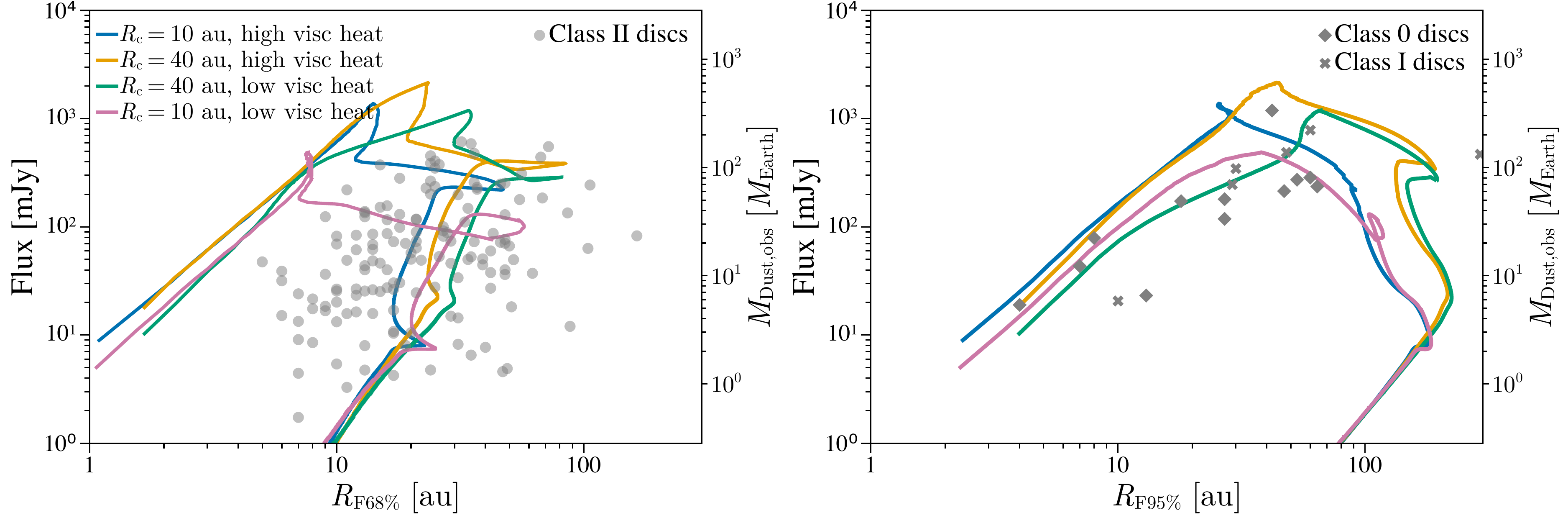}
    \caption{Disc fluxes at $\lambda=1.3$\,mm as a function of flux radius, similar to Fig. 4, in a disc created from a 1\,$M_\odot$ cloud core, comparing different combinations of high and low angular momentum, and strong or weak viscous heating. High angular momentum results in larger discs, at both the 68\% and 95\% limit, with a higher flux, while lower viscous heating reduces the flux and increases the 68\% flux radius, but has a negligible effect on the 95\% flux radius. The combination of weak viscous heating and high angular momentum clearly provide the best agreement with the 95\% radii of the class 0/I discs.}
    \label{fig:Flux_Rflux_1Ms_comp}
\end{figure*}

The grey circles in the left panel of Fig. \ref{fig:Flux_Rflux_lowvisc} show the fluxes and 68\% flux radii of class II discs taken from \cite{2016ApJ...831..125P, 2018ApJ...869L..41A, 2018ApJ...859...21A, 2019MNRAS.482..698C, 2019ApJ...872..158A, 2021AJ....162...28V}, and the grey diamonds and stars in the right panel show fluxes and 95\% flux radii of class 0 and I discs respectively, from the recent eDisk (Early Planet Formation in Embedded Disks) survey \citep{2023ApJ...951....8O} of very young stars. This sample only includes measurements where the discs are resolved. Discs with upper limits on their size were not included. The grey stars show systems that are known to be binaries. The dust disc sizes from the eDisk survey were estimated from Gaussian fitting to the continuum emission. However, not all the discs in this sample were well-fitted by a Gaussian and as such some of them might have considerable errors \citep{2023ApJ...951....8O}. The model overlaps the class II discs well on the radial drift branch, but it is unable to reproduce the very largest discs, and disc with small sizes ($\lesssim 15$\,au) and moderate fluxes ($\sim$$100$\,mJy). There is also a good match with the class 0/I disc during the branch representing the embedded disc phase (right panel of Fig. \ref{fig:Flux_Rflux_lowvisc}). In Appendix \ref{app:Orion} we show the flux-size relationship compared to a larger sample of class 0/I discs in Orion from the VANDAM (VLA/ALMA Nascent Disk and Multiplicity Survey) survey \citep{2020ApJ...890..130T}.

The continuum flux emitted from a protoplanetary disc is determined by the temperature and mass of the disc. The flux radius will thus depend both on the temperature structure and physical size of the disc. A disc with a hotter inner region will have a smaller flux radius compared to a colder one, given that other disc parameters remain the same. However, if the inner disc is optically thick, the contribution of the inner disc to the total flux will be small, and the effect of the temperature structure on the disc size will be reduced. The physical size of a disc is in part determined by the angular momentum of the cloud core it formed from. More angular momentum results in gas and dust being deposited further away from the star. With this in mind, we therefore ran model \texttt{mod-a-2-r10} to analyse how the flux and flux radius evolve with viscous heating using $\alpha = 10^{-2}$ and $R_1 = 10$\,au (equivalent cloud cores having half as much angular momentum), as used in \citetalias{2023A&A...673A.139A}. The result is shown in Fig. \ref{fig:Flux_Rflux}.

With strong viscous heating and lower angular momentum, the flux radii of the discs are smaller. Fluxes are also lower, due to the lower disc masses that are a consequence of the lower angular momentum of the cloud core. With less angular momentum, more material will fall directly onto the star rather than onto the disc. This model is able to cover the low-radius end of the class II disc better than Fig. \ref{fig:Flux_Rflux_lowvisc}. However, it struggles to reproduce the large ($\gtrsim40$\,au) class II discs. Comparing to the class 0 and I objects in the right panel, this model overestimates fluxes and underestimates radii. The overlap between model and observations happen mainly after the embedded phase of the disc has ended.

Now, we examine the individual effects of increasing the angular momentum or weakening viscous heating separately. This is illustrated in Fig. \ref{fig:Flux_Rflux_1Ms_comp}, which shows a comparison of models with different combination of high and low angular momentum budgets, and weak and strong viscous heating. Both the contribution from less efficient viscous heating (pink line)  and from the higher angular momentum (yellow line), increase the 68\% flux radii to a similar degree. However, their effects on the flux are opposite. A disc with less viscous heating is colder and therefore less bright. Higher angular momentum results not only in a larger disc, but also a more massive disc. When discs are optically thin, a more massive disc emits a higher flux. 
The 95\% flux radius (right panel) is less sensitive to the effects of viscous heating, and more so to an increase in the angular momentum. This is because the 95\% flux radius traces the faint outer regions of the disc better, and is thus more sensitive to the physical size of the disc.

In these models, we did not include photoevaporation. The inclusion of photoevaporation results in a gap opening at a radius of about 10 au after 2-3 Myr of disc evolution (see Figure \ref{fig:PPD_Evo_A23_repl}). Outside the gap a small amount of dust would be retained in a narrow ring. As the dust inside the gap radius is rapidly depleted and its contribution to the total flux vanishes, the flux radius would grow or halt at the gap radius. Continued disc evolution results in the gap radius expanding further outwards. Therefore, models with photoevaporation would result in large discs (10-100 au) with low fluxes.

\section{Synthetic population} \label{sec:res:popsynth}

\begin{figure*}[t]
    \centering
    \includegraphics[width=\hsize]{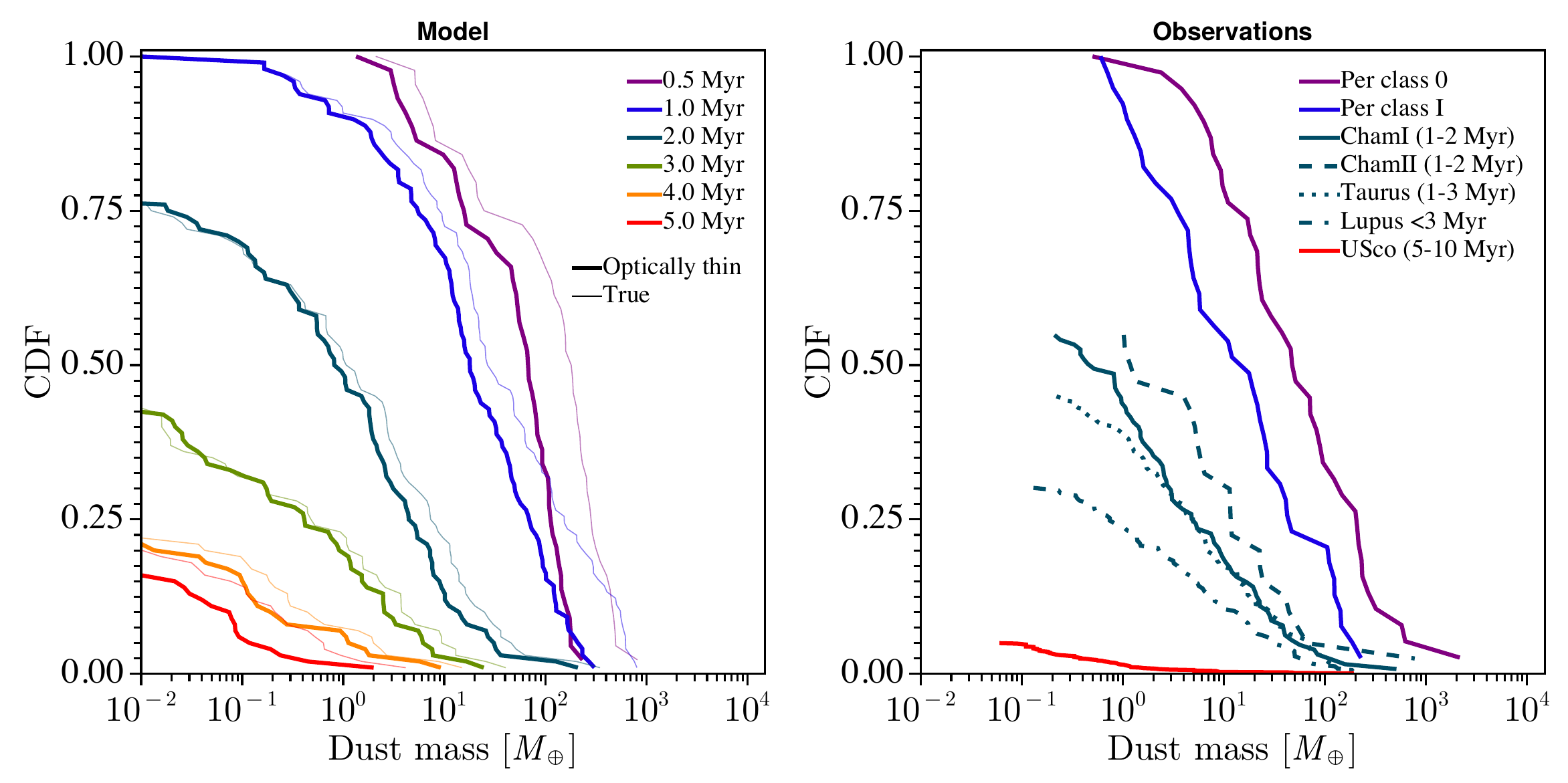}
    \caption{Cumulative distributions of dust masses in our model (left panel) and dust masses estimated from ALMA observations (right panel). Thin lines in the left panel show the true disc mass and the thick lines show the optically thin disc mass at $\lambda=1.3$\,mm. At most times, the differences between the two are small as discs are marginally or fully optically thin. The exception to this are the very young and massive discs, or old discs which have retained dust outside a photoevaporative gap.}
    \label{fig:CDF_Pic21_alpha1e-2_lowviscs}
\end{figure*}

The relation between the flux and the flux radius explored in Sect. \ref{sec:results} indicates that discs that with higher angular momentum budgets and with weaker viscous heating are better able to reproduce observed fluxes and radii. 
In particular, class 0/I discs seem to require reduced viscous heating to increase their radii for a given flux. 
With this in mind, we would like to test if these findings are also consistent with the observed evolution of the dust mass in protoplanetary discs, which decreases strongly with cluster age.
Therefore, we ran the disc population synthesis model used in  \citetalias{2023A&A...673A.139A}, but with the parameters from model \texttt{mod-a-3-r40}. 
In this way, we can investigate how the cumulative distribution of dust masses in a population of protoplanetary discs evolve under radial drift.
Unlike the models in Sect. \ref{sec:res:Flux_Rflux}, the model now includes photoevaporation in order to facilitate a comparison with the results of \citetalias{2023A&A...673A.139A}, which did include photoevaporation. We use the X-ray photoevaporation prescription from \citet{2021MNRAS.508.3611P}.
Our previous model found a generally good agreement in the dust depletion trend of our model and the observed sample (\citetalias{2023A&A...673A.139A}). However, we did not take into account the effects of optical depth on the observable dust disc mass in our model, which we can now do.

We evolve 100 discs formed from Bonnor-Ebert spheres sampled from the Kroupa IMF \citep{2001MNRAS.322..231K}. We chose to use a stellar IMF rather than the core mass function (CMF) because in a more complete treatment of star formation only $\sim$$30$‰ of the core mass is expected to end up in the star. Hence, the CMF is shifted to higher masses than the stellar IMF. Since in our model nearly all the core mass is eventually accreted onto the star, using the CMF to sample core masses would result in a distribution of stellar masses that is unrealistically massive. An important difference to the simulations in Sect. \ref{sec:results} is that we here include the \cite{2021MNRAS.508.3611P} photoevaporation model to disperse the gas disc. To emulate that all discs in a star-forming region do not form at the same time, the discs are given a random age, with a spread of 1\,Myr within the group of 100 discs.

We compare the cumulative distributions of dust masses of our model with observational measurements. We take these observed dust masses of class II discs (Chameleon, Taurus, Lupus and Upper Sco) from \cite{2013ApJ...771..129A, 2016ApJ...827..142B, 2016ApJ...831..125P, 2018ApJ...859...21A, 2021A&A...653A..46V}, and dust masses of class 0/I discs from \cite{2020A&A...640A..19T}. The cumulative distributions of the observed masses are then scaled to approximate that the disc fraction in a star forming region decreases over time as
\begin{align}
    f_\mathrm{disc} = e^{-t/\tau}, \label{eq:disc_frac}
\end{align}
where $t$ is the cluster age and $\tau$ the typical disc lifetime. We use $\tau = 2.5\, \ \mathrm{Myr}$ \citep{2009AIPC.1158....3M}, as we found in \citetalias{2023A&A...673A.139A} that this is appropriate when evolving discs at $\alpha_\nu = 10^{-2}$. 
 
The resulting cumulative distribution function is shown in Fig. \ref{fig:CDF_Pic21_alpha1e-2_lowviscs}. The thick lines in the left panel show the optically thin dust disc mass at $\lambda = 1.3$\,mm, and the thin lines show the true dust disc mass. Because the discs are mostly optically thin, the optically thin and true dust mass remain similar. A significant difference is only found for the most massive discs at a cluster age of 0.5 and 1\,Myr, or for the oldest discs where dust is retained outside a photoevaporative gap. This dust is concentrated in a narrow ring at the outer edge of the gap and the optical depth becomes very high. The right panel shows dust disc mass estimates with ALMA. The agreement between model and observations is best if the ages of the star forming regions are on the high end of the range of estimated ages.

\section{Discussion}
\label{sec:limits}
\subsection{Viscous heating in protoplanetary discs}

The presence and strength of viscous heating in protoplanetary discs is poorly constrained. A disc region where viscous accretion is the dominant source of heat results in a vertically decreasing temperature. An irradiated disc has the opposite structure, and the temperature is highest at the surface. Viscous heating could therefore be observed from the presence of spectral absorption lines, rather than emission lines. Molecular absorption lines have been found in discs around massive O-star \citep{2022ApJ...935..165B}, indicative of a viscously heated disc with a surface that is colder than the midplane. However, around T-Tauri stars the spectra are dominated by emission lines \citep{2011ApJ...733..102C}, lacking such a clear sign for the presence of viscous heating.  Viscous heating as also been inferred as the dominant heating source in the inner discs of FU Orionis objects \citep{2021A&A...646A.102L, 2024MNRAS.527.9655A}, and as a possible explanation for the increased brightness temperature in the inner region of HL Tau \citep{2024A&A...686A.298G}. 

One of the discs in the eDisk sample was recently reported showing a signature of viscous heating \cite{2024ApJ...964...24T}, with an inferred $\alpha$ value of 0.01-0.1. This disc is indicated with a red star in Figs. \ref{fig:Flux_Rflux_lowvisc} and \ref{fig:Flux_Rflux}. Interestingly, this disc sits on the formation branch in the models with  viscous heating at $\alpha_\mathrm{vh} = 10^{-3}$. In the models with viscous heating at $\alpha_\mathrm{vh} = 10^{-2}$ it is located at a position shortly after the formation branch ends. However, the stellar mass is estimated to $2.1-3.2\ M_\odot$ \citep{2023ApJ...951....8O}, which is noticeably more massive than the stars in our model. The star is also a binary, which can reduce disc sizes through tidal truncation \citep{2018MNRAS.478.2700W}. 

The physical mechanism that drives accretion remains an open question. Therefore, it is challenging to determine the strength of viscous heating. However, magnetic disc winds are likely to play an important role \citep{2013ApJ...769...76B}. In wind-driven discs, accretion heating is not a consequence of viscous dissipation, but is rather due to Joule dissipation. Simulations of the disc temperature under non-ideal magnetohydrodynamics have found Joule heating to be much less efficient than viscous heating, and it is unable to effectively heat the disc midplane under a wide range of disc conditions \citep{2019ApJ...872...98M}.

\subsection{Spectral index}

\begin{figure}
    \centering\includegraphics[width=\hsize]{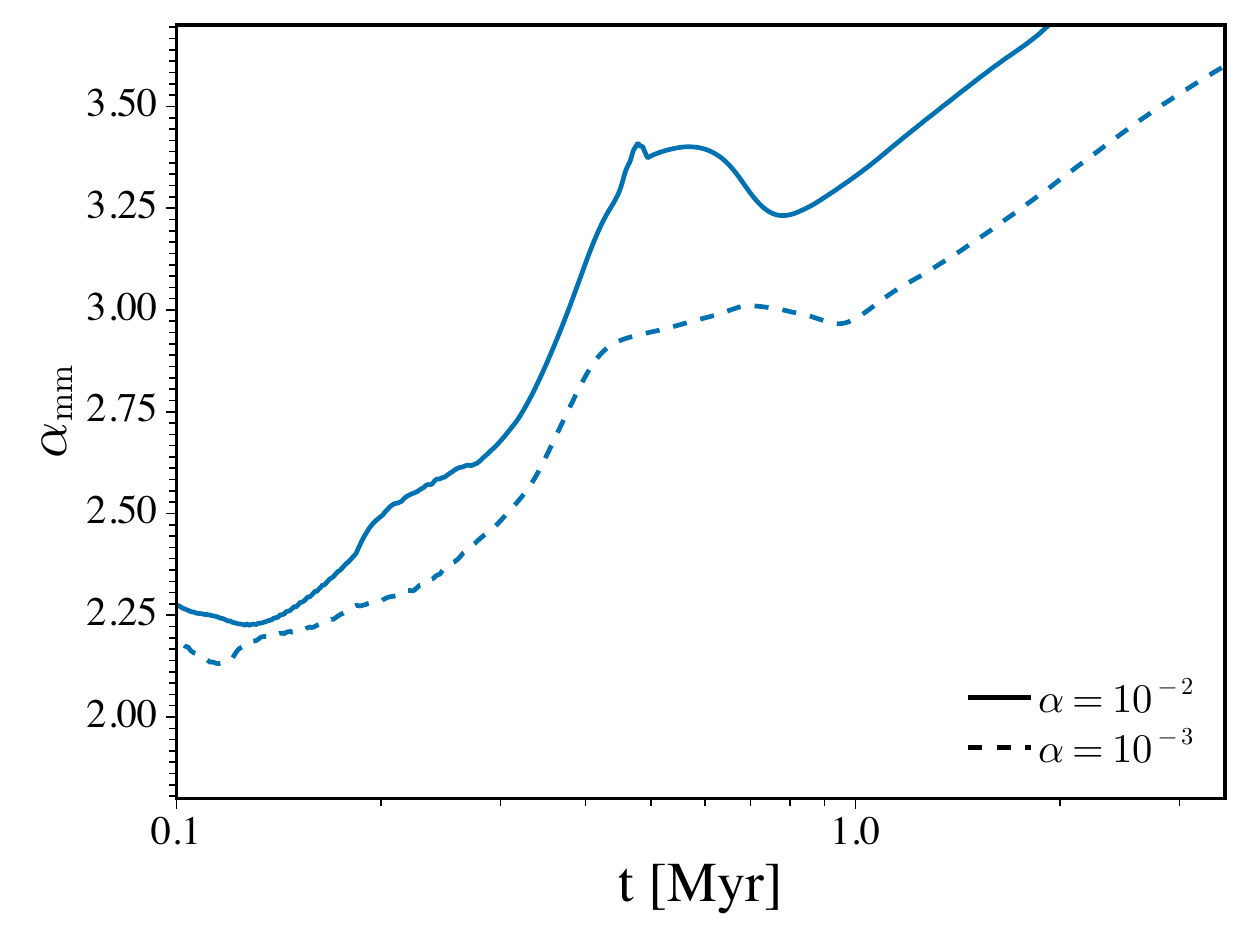}
    \caption{Spectral index of the whole disc as a function of time. The solid line shows the model with $\alpha_\nu = 10^{-2}$ and $\alpha_\mathrm{vh} = 10^{-3}$. The dashed line shows a model with $\alpha_\nu = 10^{-3}$ and $\alpha_\mathrm{vh} = 10^{-3}$. The larger particles in the $\alpha_\nu =  10^{-3}$ model results in a lower spectral index.}
    \label{fig:Spectral_indx_t}
\end{figure}

Figure \ref{fig:Spectral_indx_t} shows the millimetre spectral index of the disc as a function of time, calculated as $\alpha_\mathrm{mm} = \log_{10}\left(F_\mathrm{1.3_\mathrm{mm}}/F_\mathrm{3_\mathrm{mm}}\right)/\log_{10}\left(3/1.3\right)$. The solid line shows the nominal model with $\alpha_\nu = 10^{-2}$ and $\alpha_\mathrm{vh} = 10^{-3}$. Early on the spectral index is low $\alpha_\mathrm{mm}\sim2.25-2.5$, and as the discs evolves the spectral index increases to $\alpha_\mathrm{mm}\sim3.25$ at 1 Myr. 

The high values of $\alpha_\mathrm{mm}$ we find at 1 Myr are similar to results previous numerical studies of discs adopting a dust fragmentation velocity of 1\,m/s \citep{2021A&A...645A..70P}. This value is higher than what is typically found in observations of discs, where values are closer to $\alpha_\mathrm{mm} \approx 2-2.5$ \citep[e.g.][]{2010A&A...512A..15R}. These observed values are often interpreted as a sign of grain growth, and the higher values of $\alpha_\mathrm{mm}$ that we find could be indicative of a lack of mm-sized pebbles in the model. However, the spectral index is sensitive both to particle porosity, the opacity model used to calculate it, and to scattering in optically thick discs \citep{2014A&A...568A..42K, 2018ApJ...869L..45B, 2019ApJ...877L..18Z}. 

Lowering $\alpha_\nu$, but keeping $\alpha_\mathrm{t}$ the same, results in larger dust particles in the fragmentation-limited regime. The dashed line shows a model with $\alpha_\nu = 10^{-3}$ and $\alpha_\mathrm{vh} = 10^{-3}$. As a result, the spectral index decreases, remaining at $\alpha_\mathrm{mm}\lesssim 3$ up to 1\,Myr.

Figure \ref{fig:Spectral_indx_r_t} shows a contour plot of the spectral index in the disc as a function of time and radial position. The disc is optically thick during the first $\sim$$0.3$\,Myr and $\alpha_\mathrm{mm} \approx 2$ across most of the disc. Up to $t = 1$\,Myr, the spectral index to values ranging from $\alpha_\mathrm{mm} = 2$ in the inner disc up to $\alpha_\mathrm{mm} = 4$ in the outer discs. At later times the spectral index ranges from $\alpha_\mathrm{mm} = 3$ to $\alpha_\mathrm{mm}\gtrsim4$. The region of $\alpha_\mathrm{mm}\sim5$ traces the edge of the disc.

\begin{figure}
    \centering\includegraphics[width=\hsize]{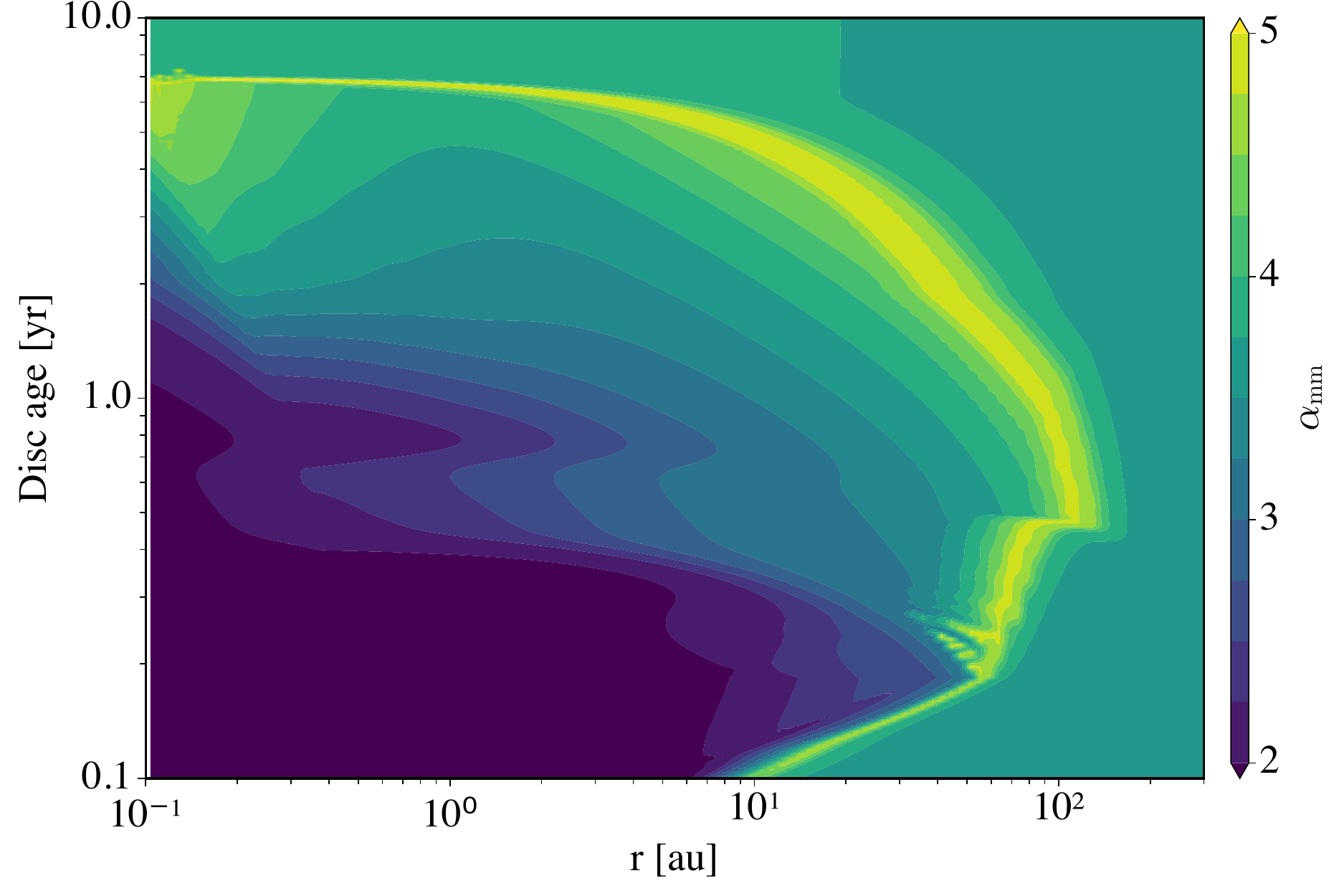}
    \caption{Spectral index across the disc as a function of radial position and time. During the early stages of its evolution ($t<0.3$ Myr) the disc is optically thick and the spectral index is 2 across almost the entire disc. When the disc has become optically thin the spectral index increases radially outwards as the particle size decreases radially.}
    \label{fig:Spectral_indx_r_t}
\end{figure}

\subsection{Model limitations}

The flux radius of a protoplanetary disc is sensitive to the diffusion of dust grains. In this work, we have not included dust diffusion, as it mainly affects the smallest dust particles and we only trace the population of the largest dust particles, which are not as sensitive to diffusion. Small dust particles can diffuse outward in the outer parts of the protoplanetary disc, increasing its size. As shown by \citep{2021A&A...645A..70P}, this can have a significant effect especially on the 90\% flux radii, increasing the radius by up to a factor of 2. The effect on the 68\% flux radius is less pronounced. Still, from the right panel Fig. \ref{fig:Flux_Rflux_lowvisc} the 95\% flux radius of our model traces the disc with the lowest observed radii the best. Hence, even if the inclusion of small diffusive dust particles would make the radii in our model larger, the agreement with the observed class 0/I discs would likely still be good.

If gas accretion in protoplanetary discs is driven by MHD winds rather than viscous evolution, this could affect the evolution of the size of the discs. In wind-driven discs, the angular momentum is removed with the wind, rather than being transported outward, as in the viscous disc model. As a consequence, the gas disc does not expand outward \citep{2013ApJ...769...76B}. However, global models have found that expansion is possible in wind-driven discs \citep{2021ApJ...922..201Y}. Because well-coupled dust particles move with the gas, a non-expanding gas disc prevents the dust disc also from expanding. In our model, this expansion of the gas disc is seen during, and shortly after, the formation of the disc (Fig. \ref{fig:Flux_Rflux_lowvisc}). For a given angular momentum budget, a wind-driven disc without expansion would, therefore, be smaller than a viscously driven disc in both gas and dust.

In our model we find that discs are mostly optically thin, and therefore using standard assumptions that go into converting continuum fluxes to dust masses give quite accurate results. However, as a number of recent studies have highlighted, disc being partially optically thick, or assuming an opacity that is not appropriate for the disc, can result in underestimating the dust mass \citep[][]{2023ASPC..534..501M}. Therefore, observational estimates of dust disc masses from converted continuum fluxes should be considered lower limits. Optically thick regions in protoplanetary discs could be present in very young and very massive discs that have yet to lose a substantial amount of dust. Another site where optically thick regions could arise are in disc with dust substructures.

Indeed, large and evolved discs frequently show dust substructures \citep{2021AJ....162...28V}. If these are dust traps, they could halt, or reduce, the degree of radial drift and accumulate pebbles \citep{2023A&A...670L...5S}, affecting the outer dust radius \citep{2019MNRAS.486L..63R}. 
Regions with high dust concentrations could be optically thick and in this way hide dust particles from observations.
At the same time, large pebbles that are retained for longer times may help in reducing the spectral index \citep{2021MNRAS.506.2804T, 2022A&A...661A..66Z, 2024ARA&A..62..157B}. 
Further exploring the effect of dust substructures would be interesting, but is challenging as their physical nature is as of yet poorly understood \citep{2023ASPC..534..423B}. Possibly they are signposts of wide-orbit giant planet formation \citep{2018ApJ...869L..42H, 2021AJ....162...28V,2024A&A...692A..45K}.

From the flux-size relationship of class 0/I discs, our models indicate that viscous heating may be operating at a reduced efficiency in protoplanetary discs. The sizes of the class 0/I discs in the eDisk sample we compare with are estimated from Gaussian fits to the disc emission. It has been found that estimating disc sizes of embedded objects from fitting a single Gaussian might overestimate the size by up to a factor of 2 \citep{2024A&A...684A..36T}. However, \cite{2022ApJ...934...95S} compared disc sizes found from fitting radiative transfer models to sizes found from Gaussian fitting. They found that Gaussian modelling of small discs likely overestimate the disc size, but for large discs the Gaussian model is quite accurate. Nevertheless, if the true disc sizes of the class 0/I sample we compare to were smaller by a factor of 2, it would not be so clear that the model discs with reduced viscous heating would match the observed sample better.

\section{Conclusions} \label{sec:conclusions}

In this paper, we studied the relationship between the flux at mm wavelengths and the characteristic flux radius of protoplanetary discs undergoing formation, expansion, and radial drift of dust. We examined the effects of the angular momentum budget of the molecular cloud core and the efficiency of viscous heating on total fluxes and flux radii. In addition to this, we calculated the optically thin dust masses in a population synthesis of protoplanetary discs to determine the effects of optical depth, and observational assumptions about disc temperature and opacity.
Our main conclusions are listed below:
\begin{enumerate}
    \item The evolution of the disc flux as a function of flux radius presents a useful tool for testing disc evolution models, as these two quantities rely on fewer assumptions to determine through observations. For our model, we find that discs with weak viscous heating and high angular momentum budgets provide the best agreement with the observed distribution of fluxes and flux radii at $\sim$$1$\,mm wavelengths, especially when compared to the class 0/I stage (Fig. \ref{fig:Flux_Rflux_lowvisc}). Discs with strong viscous heating and low angular momentum budgets result in too small sizes and too high fluxes compared to class 0/I objects (Fig. \ref{fig:Flux_Rflux}). When compared to the class II discs, neither model agrees significantly better with the observed data than the other. \\
    \item Protoplanetary dust discs evolving under radial drift are optically thin at $\sim$$1$\,mm wavelengths for the majority of their evolution. The frequently used dust mass estimates assuming optically thin discs are therefore able to trace the true mass within a factor of approximately 2 (Fig. \ref{fig:res:Mdust_obs}). During the disc formation phase, the discs are more optically thick and the dust masses are underestimated by up to a factor of approximately 3 at a wavelength of 1.3\,mm. At a wavelength of 9\,mm, the discs are optically thin during both the formation phase and subsequent disc evolution. \\
    \item Discs with large initial radii are able to sustain pebble fluxes for sufficient time to explain the observed decrease in dust disc mass with time in nearby star forming regions (Fig. \ref{fig:CDF_Pic21_alpha1e-2_lowviscs}).    
\end{enumerate}

\begin{acknowledgements}
J.A. acknowledges the Swedish Research Council grant (2018-04867, PI A.\,Johansen). A.J. acknowledges funding from the Knut and Alice Wallenberg Foundation (Wallenberg Scholar Grant 2019.0442), the Swedish Research Council (Project Grant 2018-04867), the Danish National Research Foundation (DNRF Chair Grant DNRF159), the Carlsberg Foundation (Semper Ardens: Advance grant FIRSTATMO), and the Göran Gustafsson Foundation. M.L. acknowledges ERC starting grant 101041466-EXODOSS. The National Radio Astronomy Observatory is a facility of the National Science Foundation operated under cooperative agreement by Associated Universities, Inc.

\end{acknowledgements}

\bibliographystyle{aa}
\bibliography{References}


\begin{appendix}
\FloatBarrier 
\section{Class 0/I objects in Orion} \label{app:Orion}

\begin{figure}[h]
    \centering
    \includegraphics[width=\hsize]{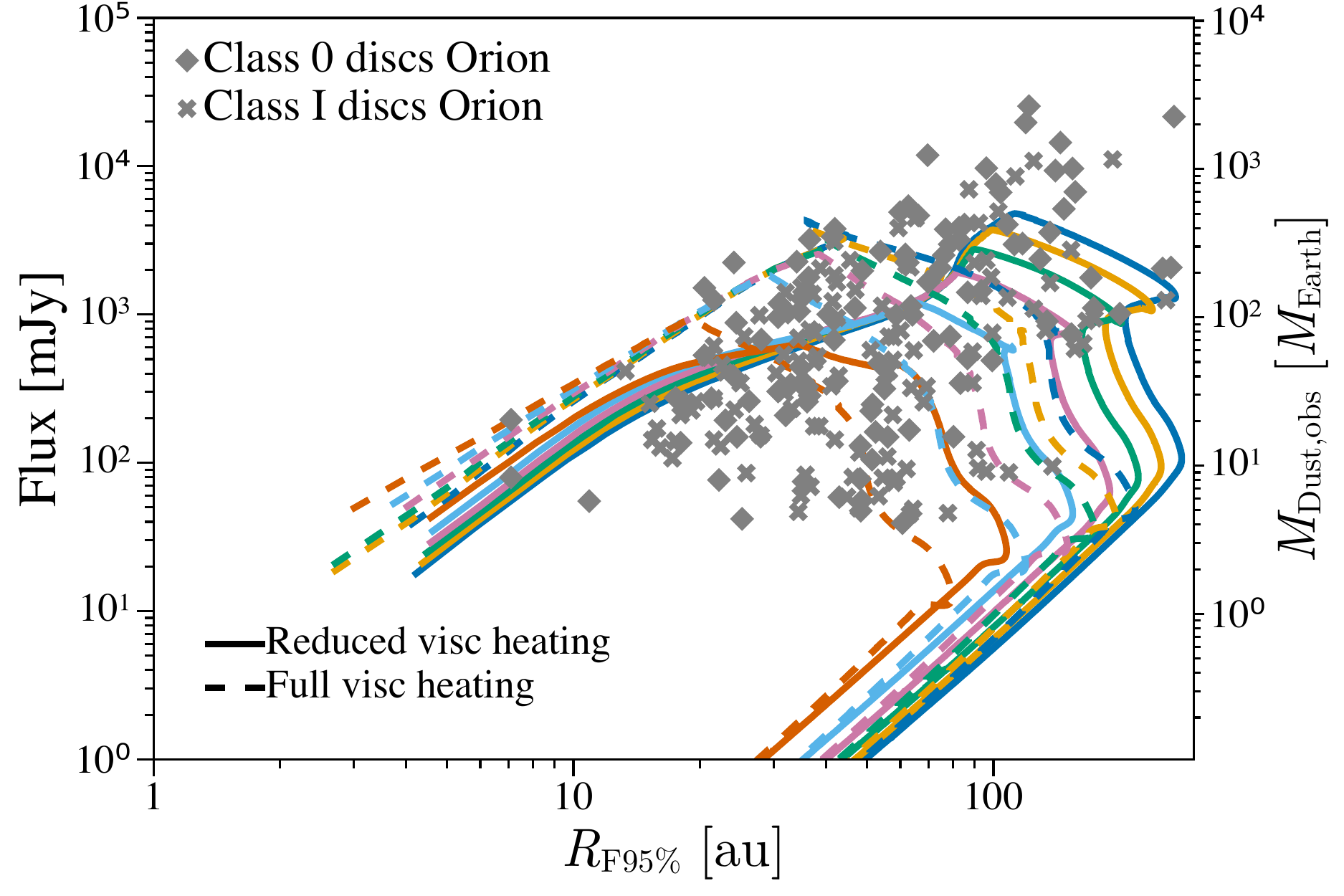}
    \caption{Continuum fluxes at $\lambda = 0.89\ \mathrm{mm}$ and sizes of our disc models compared to class 0/I discs from the Orion star forming region. Solid lines show model \texttt{mod-a-3-r40} and dashed lines model \texttt{mod-a-2-r10}. Similar to the eDisk sample shown in Fig. \ref{fig:Flux_Rflux_lowvisc} and Fig. \ref{fig:Flux_Rflux}, the model with less efficient viscous heating and higher angular momentum agrees better with the observed sample.
    }
    \label{fig:Flux_Rflux_Orion}
\end{figure}

In Sect. \ref{sec:res:Flux_Rflux} we compared the evolution of the flux-size relationship with class 0/I discs from the eDisk survey. This survey samples nearby ($d<200$\,pc) low-mass young stellar objects. This is done to remain consistent with the class II discs we also compare with, that are similarly composed of nearby low-mass star forming regions. The eDisk sample is however quite small, and may not be statistically significant. A larger sample of 230 class 0/I discs with measured fluxes and sizes exists from the VANDAM survey of Orion \citep{2020ApJ...890..130T}.

A comparison of our model to the Orion discs is shown in Fig. \ref{fig:Flux_Rflux_Orion}. The solid lines show the model with reduced efficiency of viscous heating, and the dashed lines the model with full viscous heating. As can be seen, also when comparing to the VANDAM sample, the model with reduced efficiency of viscous heating is better able to reach the fluxes and sizes found in the observed sample during phase of disc formation. The model with viscous heating at full efficiency results in too low disc sizes.

\section{Cumulative distribution of fluxes}

Instead of comparing the dust masses of protoplanetary discs in nearby star forming regions with the optically thin dust masses from our model, we can directly compare the fluxes. Figure \ref{fig:CDF_flux} shows the cumulative distribution of continuum fluxes at $\lambda = 1.3\ \mathrm{mm}$ of the population synthesis model in the top panel (solid lines) and fluxes at $\lambda = 9.0\ \mathrm{mm}$ in the bottom panel (solid lines). The dashed lines represent observed fluxes of class II discs observed at $\lambda = 1.3\ \mathrm{mm}$ in the top panel, and class 0/I discs observed at $\lambda = 9.0\ \mathrm{mm}$ in the bottom panel. For this comparison, class II discs from the Lupus, Taurus and Chameleon forming regions shown in Fig. \ref{fig:CDF_Pic21_alpha1e-2_lowviscs} were merged into one sample and the CDF was scaled to a cluster age of 2 Myr. The class 0/I discs from Perseus were also merged into one sample, and the class 0/I discs from Orion were also merged.

\vspace{0.75cm}
The class II discs at $\lambda=1.3$\,mm agree quite well with the model discs. However, the fluxes of the class 0/I discs from Perseus and Orion at $\lambda = 9$\,mm are higher than what the model produces.

\begin{figure}[t]
    \centering
    \includegraphics[width=\hsize]{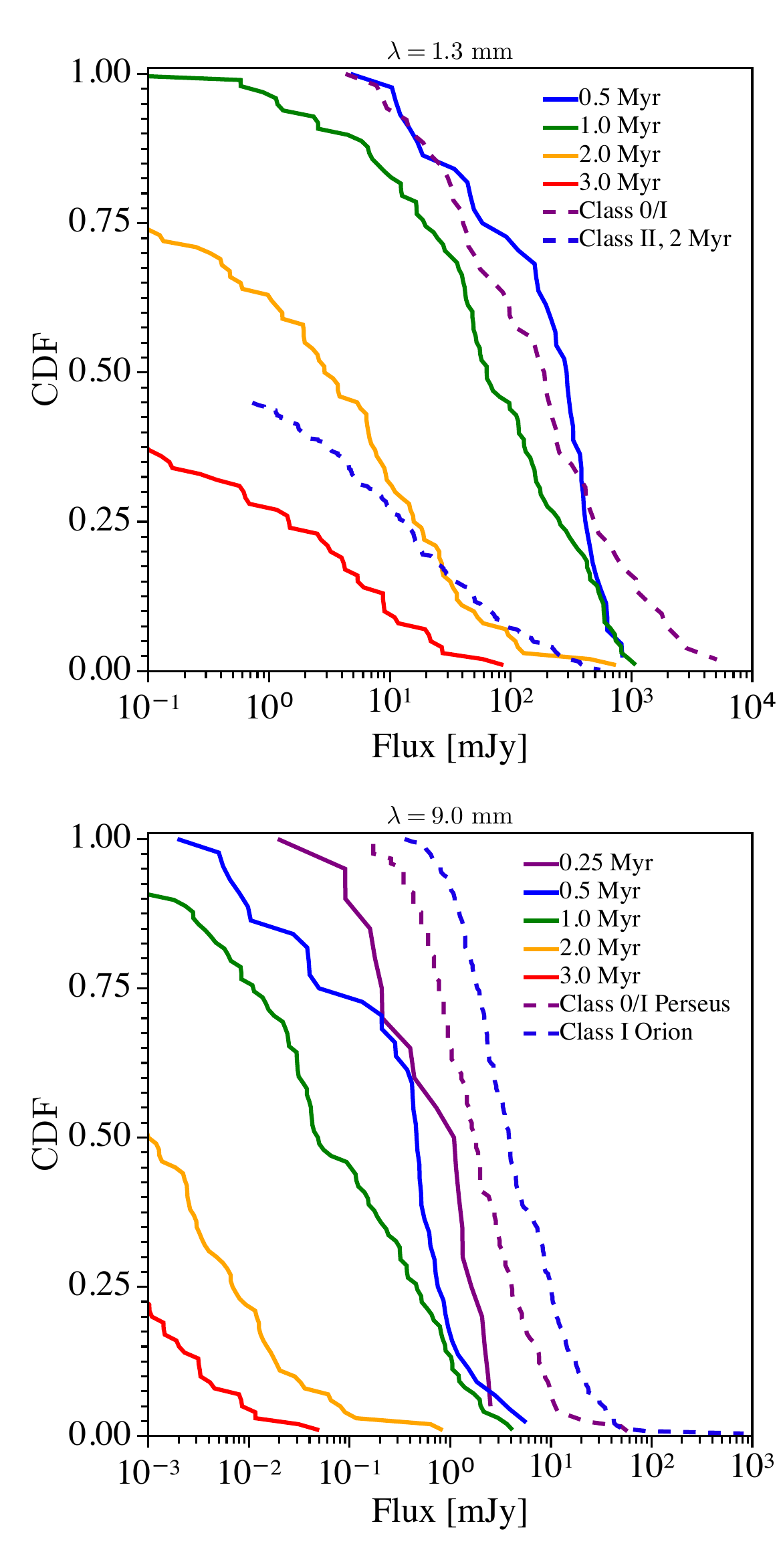}
    \caption{Cumulative distribution of continuum fluxes of the population synthesis model presented in Sect. \ref{sec:res:popsynth}. The top panel shows fluxes at a wavelength of $\lambda = 1.3 \ \mathrm{mm}$ and the bottom panel at $\lambda = 9.0 \ \mathrm{mm}$. The sample of class II discs from Taurus, Lupus and Chameleon has been merged into one and scaled to an aged of 2 Myr, and the sample of class 0/I one discs from Perseus has also been merged.
    }
    \vspace{-0.25cm}
    \label{fig:CDF_flux}
\end{figure}

\vspace{1cm}
\section{Evolution of a disc from a 1\,$M_\odot$ cloud core with photoevaporation}

In \citetalias{2023A&A...673A.139A} we presented the results of a nominal run of a disc forming from the collapse of a 1\,$M_\odot$ cloud core with a centrifugal radius of 10\,au. The disc was presented as evolving under $\alpha_\nu = 10^{-2}$ and photoevaporation from the  \cite{2021MNRAS.508.3611P} model. However, after publication we discovered a parameter error in the simulation, and that the simulation was in fact run under a lower $\alpha_\nu$. Here we present the correct results of a disc evolving at $\alpha_\nu = 10^{-2}$, with the other parameter remaining the same.

\begin{figure}[t]
    \centering
    \includegraphics[width=\hsize]{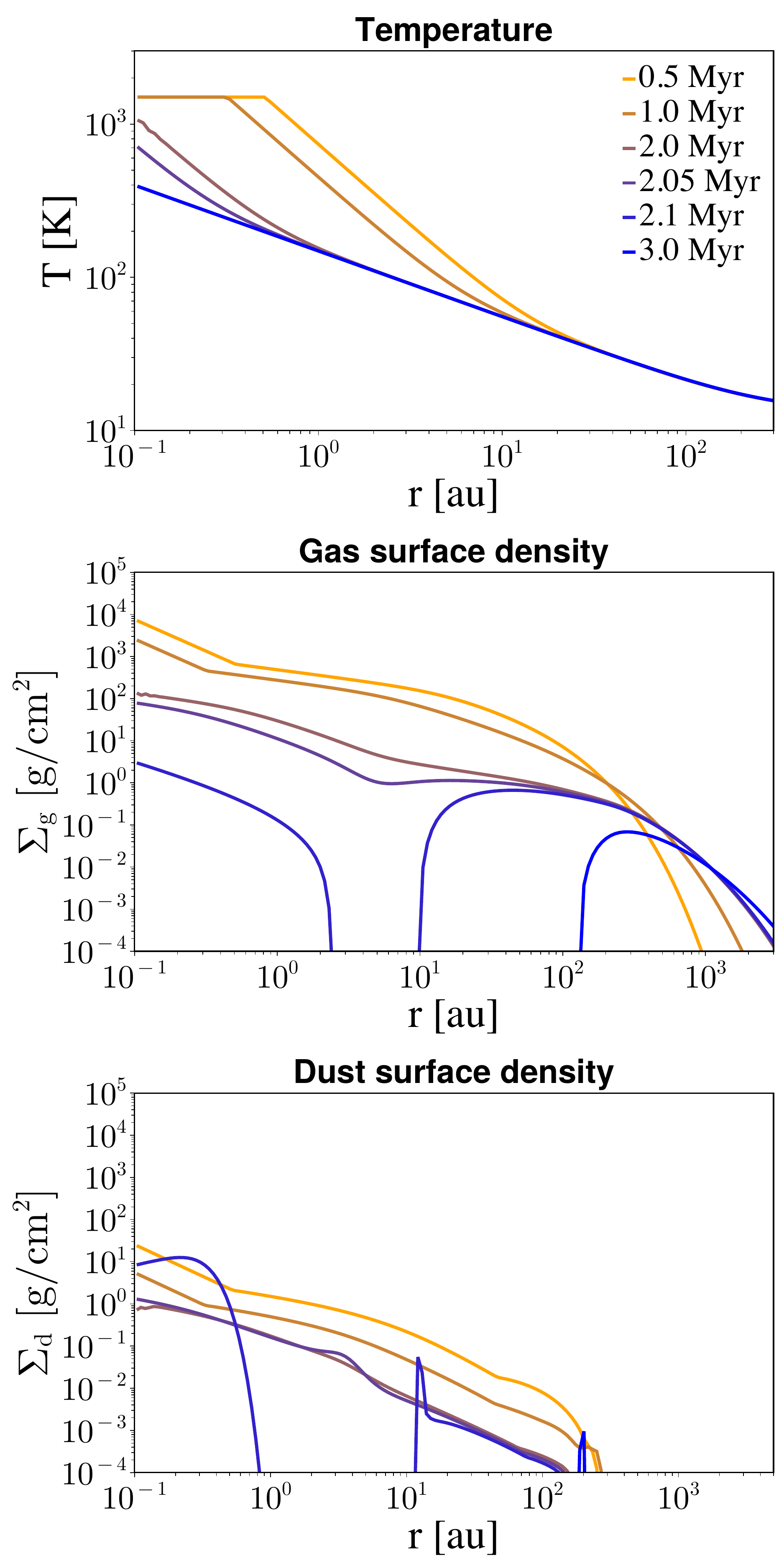}
    \caption{Temporal evolution of the disc temperature (top), gas surface density (middle), and dust surface density (bottom). Photoevaporation opens a gap in the gas after about 2\,Myr. After the gap is opened, the inner gas disc is quickly cleared by photoevaporation and viscous evolution. Outside the gaps, $\sim$$0.85\, \ M_\Earth$ of dust is retained.}
    \label{fig:PPD_Evo_A23_repl}
\end{figure}

The evolution of the disc is shown in Fig. \ref{fig:PPD_Evo_A23_repl}, which shows the temperature evolution in the top panel, the gas surface density evolution in the middle panel, and the dust surface density in the bottom panel. This figure differs from \citetalias{2023A&A...673A.139A} - after publication we discovered that Fig. 2 was produced with $\alpha_\nu = 10^{-2.7}$ as opposed to $\alpha_\nu = 10^{-2}$ as stated there. The correct results of a disc evolving at $\alpha_\nu = 10^{-2}$ is shown here, with the other parameter remaining the same. This error only affected the nominal run shown there, and not the population synthesis simulations in A23. Qualitatively, the evolution of the new results are identical to the old results. However, some quantitative results change. Chief of these is the time which it takes for the disc to evolve to a point where photoevaporation opens a gap in the disc. In the results of \citetalias{2023A&A...673A.139A} a gap opened up at $\sim$$3.7$\,Myr and in these updated results a gap opens at $\sim$$2.05$\,Myr. The amount of dust retained outside the photoevaporative gap is increased by a small amount. In \citetalias{2023A&A...673A.139A}  $\sim$$0.60\ M_\Earth$ of dust was retained outside the photoevaporative gap and in the new results $\sim$$0.85 \ M_\Earth$ are retained.

Since the nominal model in \citetalias{2023A&A...673A.139A} primarily served the purpose of qualitatively explaining the evolution of a protoplanetary disc in our model, and the qualitative behaviour of the disc remains unchanged, none of the conclusions of \citetalias{2023A&A...673A.139A} change.

\end{appendix}

\end{document}